\documentclass{mn2e}
\usepackage{amssymb}
\usepackage{ulem}
\usepackage{color}
\input epsf.sty
\newif\ifAMStwofonts
\AMStwofontstrue

\begin{document}

\title[Blandford-Znajek mechanism for low angular momentum accretion]
{On the efficiency of the Blandford-Znajek mechanism for low angular momentum relativistic accretion}

\author[Das \& Czerny]
 {Tapas K.~Das$^1$\thanks{tapas@mri.ernet.in} and B.~Czerny$^2$\thanks{bcz@camk.edu.pl}\\
 $^1$HRI, Chhatnag Road, Jhunsi, Allahabad 211 019, India\\
 $^2$Nicolaus Copernicus Astronomical Center, Bartycka 18,
      00-716 Warsaw, Poland}

\maketitle
\begin{abstract}
Blandford-Znajek (BZ) mechanism has usually been studied in the literature for accretion 
with considerably high angular momentum leading either to the formation of a cold 
Keplerian disc, or a hot and geometrically thick sub-Keplerian flow as described 
within the framework of ADAF/RIAF. However, in nearby elliptical galaxies, as well as 
for our own Galactic centre, accretion with very  low angular momentum is prevalent. 
Such quasi-spherical strongly sub-Keplerian accretion has complex dynamical features 
and can accommodate stationary shocks. In this letter, we present our calculation for the 
maximum efficiency obtainable through the BZ mechanism for complete general relativistic 
weakly rotating axisymmetric flow in the Kerr metric. Both shocked and shock free 
flow has been studied in detail for rotating and counter rotating accretion. Such study 
has never been done in the literature before. We find that the energy extraction efficiency 
is low, about 0.1\%, and increases by a factor 15 if the ram pressure is included. Such an 
efficiency is still much higher than the radiative efficiency of such optically thin flows. 
For BZ mechanism, shocked flow produces higher efficiency than the shock free solutions and 
retrograde flow provides a slightly larger value of the efficiency than that for the prograde flow.
\end{abstract}


\section{Introduction}
\label{sec:introduction}
Emission from the vicinity of black holes is usually considered as powered by accretion. However, as proposed by Blandford \& Znajek (1977), the kinetic energy of the spinning black hole may also provide another reservoir of the energy, which can be extracted with the help of the magnetic field anchored in the surrounding plasma. Such energy is believed to power jets, and is expected to be small or comparable to the nominal accretion luminosity (Ghosh \& Abramowicz 1997).

The efficiency of the aforementioned Blandford-Znajek (BZ hereafter) mechanism is determined by the properties of the medium surrounding a black hole since the magnetic strength of the magnetic field and its respective topology reflects the characteristic features of the surrounding medium.

Most of the recent studies on BZ process have been performed for flow with considerably high angular momentum 
corresponding to the circularization radius of order of hundreds gravitational radii. Accretion is then 
possible due to viscosity, and the angular momentum redistribution leads to the existence of a cold Keplerian 
or a hot sub-Keplerian disk outside the Innermost Stable Circular Orbit (ISCO), and a plunging region closer to the horizon. Within this framework, BZ mechanisms is expected to be very efficient, although the details of the efficiency calculations differ in various works. Li \& Paczynski (2000) proposed an alternating mechanism for cold disk accretion and black hole energy extraction which allows to enhance the pure accretion efficiency from 0.31 to 0.43. Hot flows are also likely to be very efficient in energy extraction (e.g. Meier 2001, De Villiers et al. 2003, McKinney 2006, Hawley \& Krolik 2006,  Nemmen et al. 2007, Wu \&
Cao 2008, Tchekhovskoy et
al. 2010). McKinney \& Gammie (2004) in their numerical simulations of a disk with net dipole moment find efficiencies up to 0.068. Garofalo (2009) suggests efficiency higher than 0.0001 if the black hole spin is higher than 0.1, and reaching 1 for classical solution, but for their flux-trapping scenario the corresponding numbers are shown to be much higher, 0.01 and 10, respectively.

Nemmen et al. (2007) and Garofalo (2009) addressed the issue of applicability of the BZ mechanism to nearby elliptical galaxies. However, the net angular momentum of such a quasi-Bondi inflow (as is believed to power such galaxies) does not have to be high. The gravitationally relaxed isotropic distribution of the old population stars
at the centre cluster of the Milky way (Ott et al. 2003), for example, may contribute the bulk of the infalling material, leading to the low angular momentum accretion flow. Currently, there is no reliable estimates of the angular momentum of the material reaching the vicinity of the black hole in such systems. The best simulations performed for Sgr A* (Cuadra et al. 2008) imply quite a large circularization radius of $5000$ gravitational radii but calculated at the inner edge of the computational domain, and most of the modeled clumpy gas is unlikely to reach the black hole. Thus the net angular momentum may be much lower. This
scenario can considerably change the maximum power available from the BZ energy extraction mechanism.

Since the axisymmetric low angular momentum accretion onto a spinning black hole shows remarkably different features in comparison to a standard Keplerian/super-Keplerian disc accretion (see, e.g., Das \& Czerny 2012, and references therein for further detail), there is a pressing need to study the energy extraction by BZ mechanism from low angular momentum rotating flow, both prograde and retro grade, to study the proper extraction mechanism of the spin energy of the massive black holes harboured by the nearby elliptical galaxies, and for our own Galactic centre black hole as well. Such mechanism may provide entirely different efficiency profile for the extracted energy, as well as of the corresponding luminosity function as compared with the other kind of galaxies where the accretion is of Keplerian in nature. Such difference may further finds
its use as an important tool to discriminate certain observational properties of various types of galaxies.

Motivated by the aforementioned arguments, we present a detail analytical scheme of
studying the efficiency of the BZ extracted energy from the general relativistic low angular momentum prograde and retrograde flow in the Kerr metric.

\section{Low angular momentum accretion with stationary shock}
\label{sec:flowmodel}
Our general relativistic low angular momentum accretion possesses dynamical features intermediate between the Bondi (Bondi 1952) type accretion and ADAF (Advection Dominated Accretion Flow, see, e.g., Narayan \& Yi 1994).
The angular momentum at the outer flow boundary is zero in
spherical flow, is locally a fraction of the Keplerian angular momentum in ADAF, and {it is of order of the Keplerian angular momentum at ISCO} for the flow model considered here.
The corresponding circularization radius is not too large in comparison to the innermost stable circular orbit (ISCO), hence the salient features of such flow in quite similar to the spherical flow at the outer region whereas significant departure in the inner region is realized through the appearance of the angular momentum barrier  and stationary shock may form as a consequence. {Thus our model, as compared to Bondi flow as well as ADAF/RIAF, however, has richer dynamical features, allowing the flow to become multi-transonic and to accommodate a standing shock}.

The detailed formulation of the equations governing the flow to be considered here and the procedure to obtain its general solutions analytically are provided in Barai, Das \& Wiita (2004) and Das \& Czerny (2012),
while the comprehensive characteristic features of such flow from the dynamical systems point of view
are available in Goswami, Khan, Ray \& Das (2006). The numerical HD and MHD simulations of such a flow were performed by Proga \& Begelman (2003a,b). Here we use analytical solutions to cover the whole parameter range.

The flow is assumed to posses considerable advective (radial three-velocity) velocity $u$ to make the infall time scale smaller than the viscous time scale in order to validate the inviscid approximation. Adiabatic accretion is considered. Two first integrals of motion, the conserved
specific energy ${\cal E}$ (the relativistic Bernoulli's constant) and the mass accretion rate ${\dot {\rm M}}$ are obtained for such flow. The geometry of the flow is determined by a local flow thickness: function of the radial distance, sound speed and various other variables -- that can be obtained by solving the general relativistic Euler equation in the vertical direction. The flow profile is essentially solved on the equatorial plane with vertically averaged values for the
thermodynamic variables, and the complete stationary accretion solution is specified by four astrophysicaly relevant parameters
$\left[{\cal E},\lambda,\gamma,a\right]$, $\lambda,\gamma$ and $a$ being the specific flow angular momentum, the adiabatic index and the black hole spin, i.e., the Kerr parameter, respectively.

The space gradient for the flow velocity $u$ as well as the sound speed $c_s$ can be cast into a set of first order autonomous dynamical systems as:
\begin{equation}
\frac{du}{dr}=
\frac{\displaystyle
\frac{2c_{s}^2}{\left(\gamma+1\right)}
 \left[ \frac{r-1}{\Delta} + \frac{2}{r} -
       \frac{v_{t}\sigma \chi}{4\psi}
 \right] -
 \frac{\chi}{2}}
{ \displaystyle{\frac{u}{\left(1-u^2\right)} -
 \frac{2c_{s}^2}{ \left(\gamma+1\right) \left(1-u^2\right) u }
 \left[ 1-\frac{u^2v_{t}\sigma}{2\psi} \right] }}
\label{eq1}
\end{equation}
and
\begin{equation}
\frac{dc_s}{dr}=
\frac{c_s\left(\gamma-1-c_s^2\right)}{1+\gamma}
\left[
\frac{\chi{\psi_a}}{4} -\frac{2}{r}
-\frac{1}{2u}\left(\frac{2+u{\psi_a}}{1-u^2}\right)\frac{du}{dr} \right],
\label{eq2}
\end{equation}
from the differential solution of the energy conservation as well as the mass conservation equation of the following form:
\begin{equation}
{\cal E} =
\left[ \frac{(\gamma -1)}{\gamma -(1+c^{2}_{s})} \right]
\sqrt{\left(\frac{1}{1-u^{2}}\right)
\left[ \frac{Ar^{2}\Delta}{A^{2}-4\lambda arA +
\lambda^{2}r^{2}(4a^{2}-r^{2}\Delta)} \right] } \label{eq3}
\end{equation}
and \begin{equation}
{\dot M}=4{\pi}{\Delta}^{\frac{1}{2}}H{\rho}\frac{u}{\sqrt{1-u^2}} ,
\label{eq4}
\end{equation}
$H$ being the local flow thickness:
\begin{equation}
H=\sqrt{\frac{2}{\gamma + 1}} r^{2} \left[ \frac{(\gamma - 1)c^{2}_{s}}
{\{\gamma - (1+c^{2}_{s})\} \{ \lambda^{2}v_t^2-a^{2}(v_{t}-1) \}}\right] ^{\frac{1}{2}} ,
\label{eq4a}
\end{equation}
where $r,{\rm and} \rho$ are
the radial distance on the equatorial plane and the flow
density. $A$ and $\Delta$ are
functions of $r$ and $a$, $v_t$ is function of $u,r,\lambda,a,A,\Delta$,
and other terms are functions of
above quantities and of various metric elements as well as their
space derivatives in the Boyer-Lindquist co-ordinate. Further details are available in Barai, Das \& Wiita (2004)
and Das \& Czerny (2012).

The fixed point analysis reveals that the flow described by the above model can have multiple (at most three) critical points, with a centre type point flanked in between two saddle type points, for certain regions of the four dimensional parameter space spanned by $\left[{\cal E},\lambda,\gamma,a\right]$. Stationary shock may form if for such multi-critical flow the general relativistic Rankine Hugoniot condition gets satisfied. For flow with three critical points, the post shock flow will finally pass through the inner sonic point while the the shock free solution passes only through the outer sonic point, as demonstrated in figure 2. of Das \& Czerny (2012).
The shock location, strength, and the
ratio of various pre- to post shock dynamical and thermodynamic variables can be obtained as a function of the Kerr parameter, which indicates how the spin of the black hole influences such quantities. Dependence of such quantities on ${\cal E},\lambda$ and $\gamma$ can also be studied as well.
\section{Calculation of efficiency for the BZ mechanism}
We start with the general formula for the BZ luminosity
\begin{equation}
L_{BZ} = {1 \over 32} {\Omega_F (\Omega_H - \Omega_F) \over \Omega_H^2} B_H^2 r_H^2a^2c,
\end{equation}
where $\Omega_F$ is the angular velocity of the zero field observers, $\Omega_H$ is the angular velocity of the black hole horizon, $r_H$ is the horizon radius, $B_H$ the normal component of the magnetic field. We calculate the efficiency of the energy extraction as
\begin{equation}
\eta = {L_{BZ} \over \dot M c^2}.
\end{equation}

For flow variables calculated along the shocked flow passing through the inner sonic point, or along the shock free flow passing through the outer sonic point of
the multi-transonic flow, or for flow through the only available addle type sonic point for the mono-transonic flow, we define the `quasi-terminal value' (QTV hereafter)
as $V_\delta$ to be the value of any accretion variable $V$ computed at a very close proximity $r_\delta$ to the event horizon where $r_\delta=r_++\delta$, $r_+=1+\left(1-a^2\right)^\frac{1}{2}$ and $\delta=0.001 \frac {GM_{BH}}{c^2}$,
$M_{BH}$ being the black hole mass to be considered. We integrate the flow upto $r_\delta$ to obtain various QTV -- $u_\delta,\rho_\delta,T_\delta,P_\delta$ for example,
where $T$ and $P$ are the flow temperature and pressure (both the gas pressure $P^{gas}_\delta$ and total pressure $P^{tot}_\delta=P^{gas}_\delta
+\rho_\delta{u^2_\delta}$ can be computed) respectively, and use such values to compute the maximum value of the
efficiency factor $\eta$ for the energy extraction corresponding to the BZ mechanism.

We assume that the normal component of the magnetic field is in equipartition with the energy density since this maximizes the magnetic field supported by the inflowing material. We also assume that $\Omega_F = 0.5 \Omega_H$ since it further maximizes the output (Phinney 1983, Reynolds et al. 2006). This assumption is not unrealistic, as shown by Komissarov (2001).

In this work, we frequently compare the value of $\eta$ for the shocked and shock free flow to understand the influence of shock on determining the value of $\eta$. To accomplish that task, for {\it same} set of values
of $\left[{\cal E},\lambda,\gamma,a\right]$, we calculate $\eta$ from QTVs computed along the supersonic flow through the inner sonic point ($\eta$ for shocked flow) and for QTVs computed along the unperturbed supersonic flow through the outer sonic point ($\eta$ for shock free
flow) and study the dependence of two such $\eta$s on $\left[{\cal E},\lambda,\gamma,a\right]$ for prograde as well as for retro-grade flow.

It is to be noted that in our work, for the first time in literature, low angular momentum accretion with shock has been studied for the counter rotating flow.

Since the stationary solutions exist only for a subset of values of $\left[{\cal E},\lambda,\gamma,a\right]$, and the solutions with shocks cover even smaller parameter space, one requires to consider a specific limited range of fixed parameters to study the variation of
$\eta$ on $\left[{\cal E},\lambda,\gamma,a\right]$ for flow with shock. All
such issues are further clarified in section \ref{sec:results} using the respective figures.

The value of $\eta$ does not depend on the adopted value of the black hole mass and the QTV of the local density beyond the Bondi radius. However, it depends on the remaining parameters: the black hole spin, asymptotic value of the Bernoulli constant, $\cal E$,  (or, equivalently, the temperature), and the polytropic index, $\gamma$.
\begin{figure}
\vskip -1.2truecm
\epsfxsize=7.5cm
\epsfbox{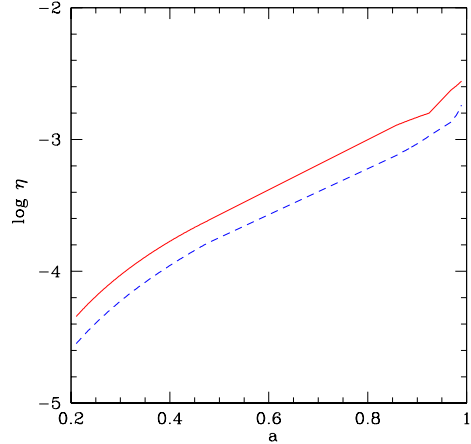}
\vskip -0.5truecm
\caption{The efficiency of energy extraction from the low angular momentum prograde accretion as a function of the black hole spin for solutions with (red continuous line) and without (blue dashed line) a shock. For the entire region of the black hole spin $a$ shown in the figure,
the angular momentum $\lambda$ varies from 2.6 (lower values of
$a$) down to 2.01 (higher values of $a$) to
ensure the existence of stationary solutions. Fixed parameters are $\gamma = 1.43$, ${\cal E} = 1.00001$.}
\label{fig:akerr}
\end{figure}
\begin{figure}
\vskip -1.5truecm
\epsfxsize=7.5cm
\epsfbox{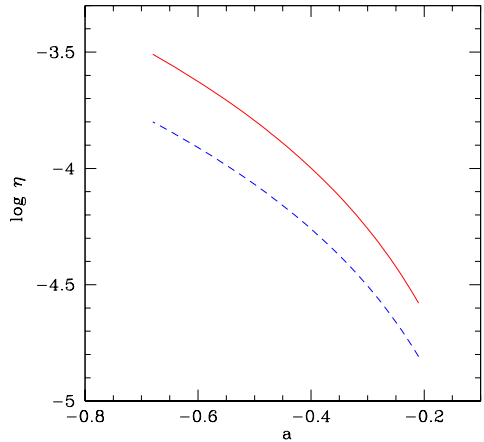}
\vskip -0.5truecm
\caption{The efficiency of energy extraction from the low angular momentum retrograde accretion as a function of the black hole spin, for solutions with (red continuous line) and without (blue dashed line) a shock for the $\lambda=3.3$, $\gamma = 1.4$, and ${\cal E} = 1.00001$.}
\label{fig:counter}
\end{figure}
\vskip -6.0truecm
\section{Results}
\label{sec:results}
We calculate the stationary solutions for a low angular momentum flow likely applicable
to Sgr A* and black holes at the centers of elliptical galaxies. Specifically, we scale
the solutions for the black hole mass and the asymptotic local density corresponding to
the  Sgr A*, $3.6 \times 10^6 M_{\odot}$ (Eckart et al. 2006), and 26 cm$^{-3}$ (Baganoff et al. 2003), respectively. For
each of the solutions, the maximum efficiency of energy extraction through
the BZ mechanism has been determined.

In figure 1 we show the dependence of $\eta$ on black hole spin $a$ for prograde flow, for both shocked (red continuous line) and shock free (blue dashed line) solutions for fixed value of $\left[{\cal E},\gamma\right]$. To cover a considerably large range for $a$
(from slowly rotating to very fast rotation for the hole), we had to take a range of values of $\lambda$ as mentioned in the figure caption, since flow characterized by a single value of $\lambda$ does not form the shock for the entire range of $a$ analyzed here.

It is obvious from the figure that the shocked flow produces higher efficiency for energy extraction through the BZ mechanism since the post shock flow is relatively hotter and denser. This is a universal feature for dependence of $\eta$ on any of initial boundary conditions corresponding to $\left[{\cal E},\lambda,\gamma,a\right]$, as is understood from the other figures as well. Similar results for the retro grade flow  has been shown in figure 2.

The dependence of $\eta$ on ${\cal E},\lambda$ and $\gamma$ is rather weak -- it  would create a finite but small width to the plot shown in Figs. 1 and 2. For a given value of
$a$,  varying $\lambda$ gives the change in log $\eta$ by less than 0.1, varying $\gamma$ gives a change less than 0.2, and varying ${\cal E}$ from 1.000001 to 1.00001 gives a change less than 0.1. Much larger values of ${\cal E}$ can increase the efficiency but it imply unrealistically high temperature at large distances from the black hole.
The value of $\eta$ correlates with ${\cal E}$ and $\gamma$  since both factors lead to higher value of the pressure for a fixed density at the outer edge, which can be easily seen in the case of Bondi flow. However, it anti-correlates with $\lambda$ for shocked case which is a manifestation of the fact that lower value of the angular momentum produces the shock closer to the event horizon and hence larger amount of energy can be extracted for such cases. $\eta$, however, correlates with $\lambda$ for shock free transonic solutions.

It is instructive to study whether the efficiency of the energy extracted through the BZ mechanism is different for the counter rotating flow. One way to accomplish that task is to study the dependence of $\eta$ on the {\it same} value (magnitude wise) of the Kerr parameter but for both the prograde and the retrograde flow, by keeping the values of $\left[{\cal E},\lambda,\gamma\right]$ constant for those two cases.
Such observation has been presented in the figure 3 for a fixed set of values of $\left[{\cal E}=1.2,\lambda=2.0,\gamma=1.6\right]$. Such high value of ${\cal E}$ may not always be realized for many astrophysical sources, however, such a value provides the transonic flow solution for the {\it entire} range of the Kerr parameter from ${a=-1}$ to $a=+1$, and hence is useful for the aforesaid comparative study. Here the flow is mono-transonic and passes
through a saddle type inner sonic point.

\begin{figure}
\epsfxsize=12.0cm
\vskip -2.5truecm
\epsfbox{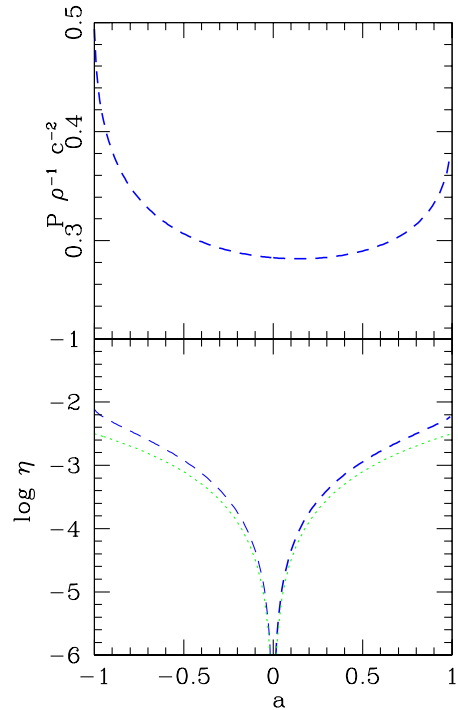}
\vskip -1.15truecm
\caption{The pressure to the density ratio close to the horizon  and the efficiency of energy extraction from the low angular momentum accretion flow for solutions without a shock covering the whole range of co-rotating and counter-rotating black holes. Fixed parameters are $\lambda = 2.0$, $\gamma = 1.6$, ${\cal E} = 1.2$. Dotted line marks the $a^2$ dependence.}
\label{fig:efipres}
\end{figure}

The solutions show the monotonic decrease of the radial position of the sonic point across the whole parameter range, $ -1 < a < 1$, changing from the value of 4.4 $GM_{BH}/c^2$ ($a = -1.0$) down
to the horizon for $a$ approaching 1. However, other flow parameters, including pressure, achieve
their minimum values at $a = 0$ and increasing with $a$ for both prograde and retrograde flows. The rise, however, is not symmetric due to the coupling with the sonic point asymmetry, as we illustrate
in Fig.~\ref{fig:efipres}. The flow efficiency is almost symmetric, with the values $7 \times 10^{-3}$ for $a = -0.99$ and $6 \times 10^{-3}$ for $a = 0.99$, since the dominant factor is the $a^2$ dependence, as shown using the dotted line,
as the flow density close to the horizon is proportional to the pressure, which simply reflects the fact that the sound speed reaches quite close to $c$ at the horizon. 

\section{Discussion}
\label{sec:discussion}
The extraction of the rotational energy from the black hole is an interesting option to power jet outflows from direct vicinity of a black hole (Blandford \& Znajek 1977). The efficiency of this mechanism is currently under thorough investigation, and it depends on the Kerr parameter as well as the properties of the inflowing material and the surrounding large scale magnetic field.

Most of the attention was payed to the matter in disk-like accretion since it this case the efficiency may be the maximum. As shown by Tchekhovskoy et al. (2011), in the extreme case of a magnetically-arrested flow this efficiency may considerably be
higher than the accretion efficiency, thus exceeding 100 per cent.

However, there are observational arguments that jets preferentially form when the accretion pattern is different from the classical Shakura-Sunyaev optically thick, geometrically thin flow (e.g. Fender et al. 2004). Such flows are generally known as RIAF (Radiatively Inefficient Accretion Flows; see Quataert 2001) and they can be divided into high accretion angular momentum flows (usually known as ADAF models) and low angular momentum accretion flows, including spherically symmetric Bondi accretion.

In this paper we analyzed for the first time the efficiency of the Blandford-Znajek mechanism in the case of low angular momentum flows. Such inflow starts at the outer boundary with an angular momentum much smaller than the local Keplerian value and proceed almost as a free fall till the angular momentum barrier becomes important. Stationary solutions with negligible viscosity exist for a range of parameters, with and without shocks. These solutions may be applicable to elliptical galaxies or to Sgr A*.
For the first ever time in the literature, we have also studied the BZ mechanism for a shocked flow, and the low angular momentum solutions for the retrograde flow. This is important because the relevance of the counter rotating flow in black hole astrophysics is being increasingly evident from recent theoretical and observational findings (Nixon et al. 2010, Dauser et al. 2011 and references therein).
The retrograde flow is more efficient by $\sim  10$ \%, so the enhancement is much lower that in the model
considered by Garofalo et al. (2007).

We find the efficiency of the energy extraction to be of the order of 0.1 \%, generally somewhat lower that in ADAF type of inflow, and much lower than in the thin disk accretion.  Narayan \& Fabian (2011) considered a model with an inflow timescale only a few times the free-fall time, due to a combination of a moderate initial angular momentum of the material and very high viscosity, and they obtained energy extraction efficiency of order of up to 2\%. In general, for a given accretion rate, the inflow velocity in the outer part of ADAF is by a factor $\alpha$ lower than the free fall, where $\alpha$ is the viscosity parameter, and the density is by a factor $1/\alpha$ higher than in low angular momentum flow in the outer parts, and this difference is only partially diminished in the innermost part, particularly in the presence of shocks. However, the expected radiative efficiency of the flow considered in our paper is also very low, emissivity calculated for a particular ex
ample of a transonic flow studied in the post Newtonian framework was of order of  $7 \times 10^{-6}$ (Mo{\'s}cibrodzka et al. 2006),
and hence the BZ energy extraction can easily dominate in such flows, as for low accretion rate ADAF solutions (e.g. Wu \& Cao 2008). The exact radiative efficiency both in ADAF and low angular momentum flow depends crucially on assumptions about the electron-ion coupling, but the radiative emissivity (synchrotron and bremsstrahlung) scale roughly quadratically with density under the same assumptions so the radiative efficiency of low angular momentum is lower roughly by a factor $1/\alpha^2$. Thus in the low angular momentum flow the BZ energy extraction more easily dominates the radiative efficiency than in ADAF.

Our low angular momentum model does not include the effect of the viscosity since the flow possesses considerable radial advection, for which the viscous time scale is fairly large compared to the
infall time scale. The stationary solutions used in this paper would not be much affected if the viscosity is included. On the other hand, large viscosity would likely broaden the parameter range for stationary solutions and the
model would become similar to ADAF.


There is also an additional mechanism which can increase the energy extraction efficiency. Narayan et al. (2003) postulated that the magnetic field can accumulate close to the black hole horizon. In this case the support for the magnetic filed can come also from the ram pressure, and from the maximum pressure at the flow, as
discussed by Narayan \& Fabian (2011). If we include the ram pressure into our flow model, the efficiency is enhanced by a factor $\sim 15$, and the model reaches the jet efficiency
of $\sim 2$ \% as required by the Chandra observations of the elliptical galaxies (Allen et al. 2006, Nemmen et al. 2007), if the black hole spin is larger than 0.9 for realistic
value of the Bernoulli constant (Fig. 1). Thus with the ram pressure included, the BZ mechanism works equally efficiently in low angular momentum model as in ADAF.
However, it never reaches the efficiency higher than 100 \% seen by Tchekhovskoy et al. (2011) due to accumulation of the magnetic field close to the horizon, so the magnetic field pressure dominates by
two orders of magnitude the ram pressure in their model.

Our consideration of the energy extraction in low angular momentum flow shows a dependence on the spin somewhat steeper than the simple $a^2$ law,
but the enhancement in comparison with this law is only by a factor 2 - 3 at the $a = 1$ limit.
It is less than the factor $\sim 100$ in the numerical results of McKinney (2005) for the BZ jet efficiency from a thick disk. It may partially reflect the difference in the dynamics, but our results may slightly underestimate the effect due to assumption of the field equipartition while McKinney (2005) and McKinney \& Gammie (2004) compute the magnetic field self-consistently.
Some observations,
however, do not confirm any obvious co relation between the jet power and the measured black hole spin in active galaxies (Broderick \& Fender 2011) or black hole binaries (Fender et al. 2010).

The role of the BZ energy extraction mechanism in jet formation is thus still an open question,
and a combination of two mechanisms is quite likely, as concluded by Foschini (2011), with an increasing role
of rotational energy extraction with decreasing object luminosity (Armitage \& Natarajan 1999), including Bondi and low angular momentum flow.

\section*{Acknowledgments}
We are grateful to Marek Sikora and Rafa{\l} Moderski for very helpful discussions, and to the anonymous referee
for providing remarks useful to improve the overall presentation of the manuscript.
This work was supported by grant NN 203 380136. Work of TKD was partially supported by astrophysics project under the XI th plan at HRI.

\ \\

\end{document}